\documentclass[sigconf]{acmart} 

\makeatletter
\def\@ACM@checkaffil{
    \if@ACM@instpresent\else
    \ClassWarningNoLine{\@classname}{No institution present for an affiliation}%
    \fi
    \if@ACM@citypresent\else
    \ClassWarningNoLine{\@classname}{No city present for an affiliation}%
    \fi
    \if@ACM@countrypresent\else
        \ClassWarningNoLine{\@classname}{No country present for an affiliation}%
    \fi
}
\makeatother

\AtBeginDocument{%
  \providecommand\BibTeX{{%
    \normalfont B\kern-0.5em{\scshape i\kern-0.25em b}\kern-0.8em\TeX}}}

\setcopyright{acmcopyright}
\copyrightyear{2023}
\acmYear{2023}
\acmDOI{XXXXXXX.XXXXXXX}

\acmConference[CHI '23]{For submission}{April 23-28,
  2023}{Hamburg, Germany}

\begin{document}

\title{Who's Thinking? A Push for Human-Centered Evaluation of LLMs using the XAI Playbook}

\author{Teresa Datta}
\affiliation{%
  \institution{Arthur}  
}
\email{teresa@arthur.ai}

\author{John P. Dickerson}
\affiliation{%
  \institution{Arthur}
  }
\email{john@arthur.ai}


\begin{abstract}
%
%
%
%

Deployed artificial intelligence (AI) often impacts humans, and there is no one-size-fits-all metric to evaluate these tools. Human-centered evaluation of AI-based systems combines quantitative and qualitative analysis and human input. It has been explored to some depth in the explainable AI (XAI) and human-computer interaction (HCI) communities. Gaps remain, but the basic understanding that humans interact with AI and accompanying explanations, and that humans’ needs---complete with their cognitive biases and quirks---should be held front and center, is accepted by the community. In this paper, we draw parallels between the relatively mature field of XAI and the rapidly evolving research boom around large language models (LLMs). Accepted evaluative metrics for LLMs are not human-centered. We argue that many of the same paths tread by the XAI community over the past decade will be retread when discussing LLMs. Specifically, we argue that humans’ tendencies---again, complete with their cognitive biases and quirks---should rest front and center when evaluating deployed LLMs.  We outline three developed focus areas of human-centered evaluation of XAI: mental models, use case utility, and cognitive engagement, and we highlight the importance of exploring each of these concepts for LLMs. Our goal is to jumpstart human-centered LLM evaluation.




\end{abstract}

\ccsdesc[500]{Human-centered computing~User models}
\ccsdesc[300]{Human-centered computing~HCI theory, concepts and models}
\ccsdesc[100]{Computing methodologies~Philosophical/theoretical foundations of artificial intelligence}
\ccsdesc[500]{Computing methodologies~Artificial intelligence}

\keywords{large language models, cognition, human-centered evaluation}

\maketitle

\section{Introduction}
In this paper, we first provide an overview of Explainable AI (XAI) and its evaluation strategies. We then provide an overview of Large Language Models (LLMs) and their current evaluation frameworks. We draw parallels between XAI and the rapidly booming field of Generative AI and LLMs, and we argue that three focus areas of human-centered evaluation of XAI must also be applied to LLM evaluation. 

We encourage work that:
\begin{itemize}
    \item seeks to understand the \textbf{mental models} of LLMs that everyday users hold
    \item evaluates the utility of LLMs via \textbf{use case} based user studies
    \item documents how \textbf{cognitively engaged} users are with LLM outputs and how this affects user and system behavior
\end{itemize}

 These are  vital for evaluating how LLMs meet real human needs (with their cognitive quirks) and work within real human environments.  We provide a brief primer on XAI and LLMs in the succeeding sections, then support our case that the lessons learned in XAI research may largely transfer to those in the more nascent LLM space.

\subsection{Explainable AI}
Defining explainability and the related notion of interpretability for machine learning (ML) models is not black and white, and is a subject of ongoing discussion.\footnote{We make no normative judgments with respect to the ongoing debate regarding explainability versus interpretability, and redirect the interested reader to the work of \cite{rudin2019stop, Rigorous_Science,survey} for a primer on that discussion.  Rather, we acknowledge that XAI---for all of its potential pitfalls and flaws---has found a home in production use of models in industry, thus spurring on a wide variety of research programs covering usability, efficacy, and validity.  Indeed, the existence of that increasingly-mature decade-long body of research motivates this paper, where we predict a similar pattern of disagreeing lines of analysis maturing in parallel with respect to LLMs.} For the purposes of our discussion, we focus on the most common type of model transparency seen in industry: post-hoc explanations of black box models. Post-hoc explainers often rely on only a trained model’s inputs and outputs to identify patterns in how the model makes decisions. These methods aim to unlock transparency in order to allow stakeholders to understand the decision-making process of models to improve trust and mitigate downstream harms~\cite{survey, Rigorous_Science}. 

One key consideration for XAI is that there is no such thing as a ground truth for explanations. While classic supervised ML  problems have ground truth labels, there is no exactly-correct explanation for how a black box produces outputs, for all but the simplest models. This lack of ground truth means that classic accuracy-based metrics cannot be measured for XAI. Instead, a variety of measurable, quantitative properties have been developed to evaluate how ``good'' an XAI method is. These properties~\cite{Whatmakesagoodexplanation} include: \textit{fidelity/faithfulness} (how consistent are explanations to the underlying black box), \textit{robustness} (how consistent are explanations to infinitesimally small changes in the input), \textit{compactness} (how measurably complex are the explanations), and \textit{homogeneity} (how do the qualities of the explanations differ across subgroups). These different properties target different notions of what a practitioner might want from an explanation, and thus can sometimes be in trade-off with one another~\cite{Whatmakesagoodexplanation}. Further, these measurable properties largely only explore the technical validity of the produced explanations. It is vital to acknowledge that there are other areas of consideration beyond these mathematical and technical formulations. 

In practice, practitioners often use XAI as an assistive tool for accomplishing some downstream task---model debugging, generating hypotheses, ensuring compliance, etc~\cite[see., e.g.][]{liao2022connecting}. This means that the context of XAI use involves some human using an explainer as a piece of evidence to make some decision. Thus, we need to consider how practitioners use, receive, and comprehend outputted explanations. This is especially true because data scientists are humans, and thus susceptible to cognitive biases such as confirmation bias. Scenarios have been identified where explanations are found to be misleading and lead to misplaced trust in faulty models~\cite{interpretinginterpretability,Chan20:Artificial}. In human-AI teams in the medical domain, studies have shown clinicians making more errors than usual when the AI system produces incorrect predictions~\cite{Jacobs2021}.  \textbf{Human-centered evaluation} of XAI focuses on defining the goals of XAI to meet real human needs and to work within real human environments. This means acknowledging the lack of perfect rationality in human decision-making.

\subsection{Large Language Models}

In the last year, Large Language Models (LLMs) have exploded in popularity---from research, to industry, to perhaps most importantly, public awareness and accessibility. These unsupervised autoregressive models predict next tokens (character, word, or string) based on past context~\cite{cognitive_LLM}. They can generate text for a variety of traditional generative tasks including summarization~\cite{zhang2023benchmarking}, dialogue~\cite{openai_2022, bard}, and more. Language models have grown increasingly \textit{larger} under pressures to increase performance. Two common identifiers of how ``large'' a language model is are its number of parameters and the size of the dataset it was trained on~\citep[see, e.g.,][for in-depth coverage]{Stochastic_Parrots}. Using two then-state-of-the-art (SotA) models as examples, in 2019 BERT~\citep{bert} featured 
$O(10^{8})$
parameters and was trained on a dataset size of 16 GB. In 2022, only 3 years later, GPT-3 \cite{openai_2022} was released, featuring 
$O(10^{11})$  
parameters and a training dataset size of 570 GB.  We, again, make no normative statements regarding the ``scale is all you need'' camp; rather, we note these are broadly acknowledged metrics in the community. 

Current LLM evaluation mechanisms include quantitative metrics measuring notions of \textit{accuracy} (how similar are the generated outputs to the expected outputs), \textit{robustness} (how resilient is the model to transformations of the input), \textit{calibration} (how meaningful are the generated probabilities in respect to uncertainty), \textit{efficiency} (what are the energy, carbon, and time costs for training and inference) and more~\cite{liang2022holistic}. Some also go beyond singular training-time loss objectives and implement reinforcement learning using human feedback in the loop~\cite{lambert2022illustrating}. 

A variety of potential harms and failure modes for LLMs have also been identified. There are substantial environmental costs associated with the volume of computational power required for training and inference~\citep{Stochastic_Parrots,Schwartz20:Green}. There are also concerns of LLMs propagating unfairness or bias. Counterfactual fairness~\citep{counterfactual} examines how perturbing the demographic signals of existing test examples can change the performance of the model (e.g. ``He worked at the local hospital'' versus ``She worked at the local hospital''). Fairness can also be evaluated via performance disparities between demographic subgroups. There are also other concerning forms of biases such as stereotypical associations, erasure, and over-representation in the semantics of its output~\cite{mattern2022understanding, liang2022holistic}. Finally, LLMs have been shown to produce toxic outputs. Toxicity in this context refers to hateful, violent, or offensive text~\cite{https://doi.org/10.48550/arxiv.2206.08325}, and has been shown to result even when the text prompt input is not itself toxic~\cite{https://doi.org/10.48550/arxiv.2009.11462}. 

LLMs often suffer from factual errors---they can ``hallucinate'' information~\cite{SteinhardtTalk} by providing very confident-sounding but entirely false responses. Chatbot LLMs have also been found to engage in disturbingly emotional personal conversations when session lengths are not limited~\citep{roose_2023}. There are also privacy concerns- work has shown that LLMs are susceptible to training data leakage under adversarial attack~\citep{carlini2021extracting}.

Although many concerning issues of LLMs have been identified, these largely only reflect distributional or instance-wise properties of their outputs. Zooming out on a human-centered paradigm, the ML community has not yet devoted efforts to consider how the presence (and potential ubiquity) of this new technology may fundamentally affect society and the ways that individual humans make decisions. We hope this work jumpstarts efforts to address this gap.

\section{Parallels between XAI and LLMs}
When examining LLMs within a human-centered paradigm, we notice that like XAI outputs, LLM outputs are often meant for some downstream decision or task---what email to send to your client, what quick summary of an important document you will read, what answer is provided for a pertinent question. These use cases can range from the high-stakes to the everyday, but align in that the outputs of the AI system, the XAI attachment to a traditional ML system or an LLM, are not the human's end goal.  Rather, they are a tool to help someone accomplish some other task, and so we must evaluate how these AI outputs help or hurt human interests. Another key similarity between XAI and LLMs is that they are both inherently open-ended systems, so there is typically no ground truth. Although recent work has argued for moving past an accuracy-first ML paradigm~\cite{10.1145/3442188.3445901}, XAI and LLMs force this to be the case with the lack of ground truth labels. 

For these reasons, frameworks of human-centered qualitative evaluation must be employed. For XAI, this is a developed field of study with an emphasis on a multitude of aspects of cognition. For LLMs, this is not the case and these qualitative aspects of evaluation have not yet been explored. \textbf{We argue that the areas of human interaction and cognition that have been dominant in XAI and HCI communities need to be held front and center for LLM research.} We need to understand how users make decisions about whether to utilize the outputs of LLMs, the mental models that users have of these technologies, whether LLM outputs are actually helpful in downstream tasks, and how much users cognitively engage with the outputs to verify their correctness and lack of harm. We dive into these three areas of concern in the following section.


\section{Framework of qualitative evaluation}
\subsection{Mental Model Matching}
A user’s mental model~\citep{design_of_everyday_things} of a technology is their internal understanding of how a technology works. For instance, maybe your mental model of a crosswalk is that pushing the crosswalk button will cause the walk signal to appear more quickly. However, for many cities, that button does not actually do anything~\cite{buttons}. This is an example of a user's mental model not aligning with a technology's true model. People rely heavily on their mental models of technology to make decisions. It has been found that XAI stakeholders use their mental models of XAI to decide when to use the technology~\cite{cabrera2023improving}, to evaluate how much to trust the outputted explanations~\cite{cabrera2023improving, he2022walking, facilitate}, and to make sense of any results~\cite{he2022walking, kaur2022sensible}. 

These personal mental models are formulated from a user's perceptions and interactions with the technology and how they believe the system works. While ML practitioners may have had access to specialized training on how LLMs work, this is decidedly \textit{not} the case for the vast majority of the general population. We cannot assume that everybody will have the same understanding of how a technology works as we do. To our knowledge, no work has explored the mental models the general public holds for any foundation model, let alone LLMs. How a general user believes an LLM to work may be very different from how it actually works, and this mismatch can be dangerous. It is not difficult to imagine frightening scenarios where users anthropomorphize or deify an LLM chatbot, understanding it to be a ``magical'' source of ground truth. This could very quickly lead to conspiracy theories and the legitimization of disinformation campaigns~\citep[see, e.g.,][]{hsu_thompson_2023}. It is important to consider if this is an issue of messaging and education---informing the public via AI literacy---or of regulators---to implement policies that force the algorithm providers to provide accurate, comprehensible warning labels on the limitations of their technology.


\subsection{Use Case Utility}

As previously discussed, XAI and LLMs are often tools for accomplishing some other goal. The term \textit{use case} in the XAI literature~\cite{liao2022connecting} refers to a specific usage scenario and its associated downstream task or end goal. It has been found in the XAI literature that although it might be easy to assume that an explanation will be helpful for a user accomplishing a task like model debugging or model understanding, this is not necessarily the case~\cite{interpretinginterpretability}. When the performance of that downstream task is measured, the presence of explanations can sometimes have no effect, or can even have a negative effect on performance, especially if the XAI is faulty~\cite{adebayo2020debugging, Jacobs2021}. Very limited work has explored the utility of LLMs in use-case--specified user studies, but a user study on Microsoft/Github's Copilot~\cite{github}, an LLM-based code generation tool, found that it ``did not necessarily improve the task completion time or success rate''~\cite{vaithilingam2022expectation}.

In response to these findings, the XAI community has pushed for a context-based approach when building explanation methods~\citep{liao2022connecting, beckh2021explainable}. Advocates push for first identifying a specific use case, then understanding the types of transparency useful and relevant for each stakeholder in that context, then designing the explanation method with these learnings held front and center, and finally evaluating how helpful the explanation was for specific tasks through practitioner user studies. In scenarios where the users are domain experts (e.g. medical treatment), this may involve long term collaboration to understand domain requirements and preferences. 

There are also parallels with validity and alignment theory. A system is \textit{valid} if it does what it purports to do. If it doesn't, it is likely because there are issues of \textit{alignment} between the intent of the system builder and the property the algorithm optimizes \cite{doi:10.1146/annurev-statistics-042720-125902}. Some work has flagged that it is irresponsible to publicly market LLMs for tasks they are not optimized for and were not designed for \cite{johnson_iziev_2022, Stochastic_Parrots}. For instance, dialogue LLM tools generate probabilistic outputs, not search results. Future works in this space could examine context-based design approaches and target alignment.

\subsection{Cognitive Engagement}

Cognitive effort is a form of labor, and unsurprisingly, people tend to favor less demanding forms of cognition and other mental shortcuts~\cite{Garbarino1997, tversky1974judgment}. As an example, when asked to ``agree'' to a user agreement when signing up for a new platform, you are probably more likely to check the box than to cognitively engage with the language of the agreement. 

Unfortunately, this human tendency can lead to unintended or dangerous outcomes because humans are susceptible to a wide variety of cognitive biases such as confirmation bias. Confirmation bias~\cite{Nickerson1998} refers to the interpreting of new evidence in ways that confirm one's existing beliefs. For XAI, this manifests as practitioners only superficially examining explanations instead of digging deeply, leading to over-trust, misuse, and a lack of accurate understanding of the outputs~\cite{interpretinginterpretability}. This can be dangerous when it results in the over-confident deployment of a faulty or biased model. Further, user's need to reach a certain level of cognitive engagement before being able to learn or gain knowledge from the AI interaction \cite{gajos2022people}. The XAI and HCI communities have deeply, albeit not exhaustively, explored these phenomena and the difficulties of measuring it accurately \cite{interpretinginterpretability, Bucinca_2020}. Those communities have also identified mitigation strategies, primarily through cognitive forcing strategies~\citep{buccinca2021trust, croskerry2003cognitive, lambe2016dual}. Forcing users to cognitively engage through some small task \textit{before} showing a system's output yielded the highest performance in a comparative study~\cite{green2019principles}. This is not new or specific to XAI. Train conductors in Japan famously point and call out important information on their journeys---a cognitive forcing method which has reduced human errors by nearly 85\%~\cite{r_2022}. 

Issues of cognitive engagement should be held front and center for researchers of LLMs. Because of their massive scale and public accessibility, LLMs may quickly become ubiquitous in all aspects of daily life. Realistically, how much will users actually cognitively engage with the magnitude of generated outputs to ensure that they are correct and aligned with their intentions? Consider an LLM-generated email: how often and how deeply will a user review that generated email before sending it? What if it's not just one email, what if it's every email? Will they always catch when the generated output says something incorrect, or worse, inappropriate. Furthermore, our attention spans have decreased dramatically with the increase in digital stimulation\cite{mark_2023}. The potential of cognitive forcing strategies as a potential route for responsible LLM use should be explored.

Another aspect of concern is that LLM outputs often \textit{sound} very confident, even if what they are saying is hallucinated~\cite{SteinhardtTalk}. When the user inquires about the incorrectness, they also have a documented tendency to argue that the user is wrong and that their response is correct. In fact, some have called LLMs "mansplaining as a service" \cite{mansplaining}. This can make it more difficult for humans to implement cognitive checks on LLM outputs. While some recent LLM work has outlined categories of failure modes for LLMs based on the types of cognitive biases use~\cite{cognitive_LLM}, we push for greater work in this field. We believe that there are unknown consequences of offloading our cognitive load onto these agents.


\section{why is this important}
While there are similarities between XAI and LLMs, there is at least one key difference---the potential scale of LLM usage is massive as a fundamentally lower-level ML-based building block than XAI. Their levels of performance have resulted in widespread public interest in practically all sectors of society. ChatGPT~\cite{openai_2022} set historic records for its customer growth, with over 100 million users in its first 2 months~\cite{guardian_2023}. Unlike XAI, whose users are largely technical practitioners \cite{umang} focusing on one pipeline, LLMs are often designed for public use and are meant to help with a wide variety of tasks. People can repeatedly integrate LLMs into their daily lives in many ways---to decide what to eat for breakfast, to write the responses to emails left unanswered from yesterday, to develop the sales pitch they have to present mid-morning, to generate a funny poem during a break from work, etc. The ability to influence or make decisions is a form of inherent power, and offloading cognition onto AI agents must first be met with caution. Other works foreshadow potential existential threats that larger language models and the trajectory towards Artificial General Intelligence (AGI) can cause~\cite{youtube_2023}. 

The consequences of not having a qualitative understanding of how humans interact with LLM outputs is grave. We urge for immediate efforts to address these gaps, and make a few suggestions for future works, taking inspiration from previous XAI research routes. We encourage work that seeks to understand the mental models of LLMs that everyday users hold, that evaluates the utility of LLMs via use case based user studies, and that documents how cognitive engagement affects user behavior. It is dangerous to continue developing and making available larger and larger language models without a proper understanding of how humans will (or will not) cognitively engage with its outputs.

\section{Conclusion}
In this paper, we push to jumpstart the human-centered evaluation of Large Language Models. We outline similarities between the more mature field of XAI and the nascent research boom of LLMs. We outline three focus areas of consideration- Mental Model Matching, Use Case Utility, and Cognitive Engagement. These all center on understanding real human needs and their cognitive quirks, the human-AI interaction effects on decision making tasks, and the potential dangers of offloading cognitive labor to these agents.




\bibliographystyle{ACM-Reference-Format}
\bibliography{sample-base}

\appendix

\end{document}